# PCR效率振荡与金纳米粒子浓度的光量子机理[*]

方欢欢 [1)] 陈永聪 [1)†] 刘泽飞 [1)] 朱晓梅 [2)] 敖平 [3)‡]

1)（上海大学物理系,上海定量生命科学国际研究中心,上海 200444）

2)（上海理工大学，光电计算机工程学院，上海 200093）

3)（四川大学，生物医学工程学院，成都 610065）

## 摘 要

纳米材料在聚合酶链式反应（PCR）技术中的广泛应用为改进生物医学领域的检测方法开辟了新途径。最新的实验（Chem. Eur. J. 2023, e202203513）揭示了PCR效率与pM区金纳米粒子浓度间的振荡行为，其产生或与带电胶体粒子间的长程库仑作用和纳米颗粒电子态的量子尺寸效应相关联。通过蒙特卡洛模拟，我们发现溶液中金纳米粒子的径向分布函数随着电荷的增加逐步呈现出峰值特征，从而引发光子在溶液中瑞利散射的相干行为，影响PCR链式反应过程中释放光子的再利用效率。研究表明振荡周期与下游反应光子的波长吻合，同时其能量与金纳米粒子费米能级附近的能级宽度相匹配。而后者可以吸收并储存于其内部的电子态，通过再释放过程可促进PCR上游反应进程，并由电子的玻尔兹曼分布弥补所需能量的不足部分。本工作有望推动PCR特有的精确探测手段在量子生物科技领域的应用。

**关键词**：金纳米粒子，聚合酶链式反应，长程库仑作用，瑞利相干散射，量子尺寸效应

† 通讯作者. E-mail: chenyongcong@shu.edu.cn
  第一作者. E-mail: 2229537512@shu.edu.cn

# 1 引言

聚合酶链式反应（polymerase chain reaction, PCR）是一种用于扩增特定 DNA 片段的分子生物学技术，极微量的 DNA 片段在该过程中以指数级扩增[1-4]。PCR 最常以定量 PCR（quantitative PCR, qPCR）的形式展开，这允许在实验运行期间更早地检测到扩增产物，并通过循环阈值（Ct 值）对感染单位进行定量[5]。从样本采集到给出检测结果通常需要 6~48 h，无法实现实时快速检测。另一方面，纳米材料与较大尺度物质相比，展现出不同的理化性质和光学性质。半导体纳米晶体（量子点）由于纳米尺度上的量子力学效应在太阳能利用和生物医学成像[6]方面有着重要应用。大量研究发现，引入金纳米粒子（Au Nanoparticle, AuNP）[7, 8]、$C_{60}$[9]、β-环糊精包被的铂纳米粒子[10]、石墨烯[11]等材料均可以不同程度地提高 PCR 效率。部分研究者认为，其或可归因于纳米粒子的高导热性能[8]以及与单链 DNA（single-stranded DNA, ssDNA）的有效结合两个方面[12, 13]。

2021 年，上海理工大学宋波课题组在 qPCR 实验中掺入了低浓度带负电荷的 AuNPs（AuNP-qPCR）[14]。结果显示，DNA 复制速率与 AuNP 浓度之间存在振荡关系。他们分析了三磷酸脱氧核苷酸（deoxynucleotide triphosphates, dNTP）水解过程中的成键和断键能量，提出其释能过程除了遵循传统的放热反应，还可能以生物光子形式释放部分能量。2023 年，复旦大学顾正龙等人基于优化后的 qPCR 系统真实探测到 DNA 复制过程中释放的中红外光子（mid-infrared, MIR），并进一步明确了在低 AuNP 浓度区实验扩增效率随着粒子间距

离增加而出现周期性波动[15]。

根据实验结果，文献[14]和[15]推断观察到的实验效率振荡现象归因于光学效应的参与，认为 AuNPs 是通过再吸收反应过程中释放的生物光子来影响 PCR 的效率。将 dNTP 的水解反应分解为四个子反应后，发现子反应 IV 释放的生物光子和子反应 I 断键所需要的光子相吻合。采用唯象分析估算了在效率振荡波峰时 AuNP 的间距与该光子波长的匹配度。猜测光子波长与特定浓度下相邻 AuNPs 形成的准一维腔的长度之间存在相关性时，实现了极化子介导的有效能量传递。

但是该唯象分析存在确定性问题，在解释实验本身时仍有不少的疑惑，尤其是生物光子在相邻 AuNP 间形成准一维驻波的图像。由于波长远大于纳米颗粒的尺寸，这在波动描述上并不成立。为了更深入地探索 AuNP-qPCR 效率提高的微观机制，本研究从理论上定量重构观测到的光学振荡现象。我们通过使用蒙特卡洛技术模拟溶液中 AuNPs 的相对平衡状态，采用瑞利散射模型得到表征实验效率的特定参考点相对光强，将理论预测与实验结果进行拟合对比。同时，本文探究了粒子在纳米尺度上的量子效应，用三维谐振子模型和无限深方势阱模型估算 AuNP 费米能级附近的能级宽度，评估其与生物光子的相互作用是否能弥补一定的能量缺损，从而对观察到的实验效率增加有所贡献。本工作旨在完善该实验现象的理论框架，为后续进一步研究奠定良好的基础。

## 2 模拟与计算

### 2.1 AuNP 的模拟

我们研究的重点是分散在溶液中的胶体 AuNP 混合物。将溶剂（近似为水）视为具有恒定介电常数的连续介质，使用长程库仑力描述 AuNP 间的相互作用。

纳米粒子的ζ电位$U(r)$约为$-20$ mV[16, 17]，表示斯特恩层和扩散层之间滑动面上的平均静电势[18]。根据文献显示，对应于 25-50 mV 的表面电位，可推算几百个有效电荷的存在[19]。由于粒子运动和扩散层动力学等因素的影响，精准确定电荷数量存在困难。

本文使用蒙特卡洛技术分析溶液中胶体粒子的分布，该方法被广泛应用于具有复杂相互作用的系统[20-22]。选择半径为 R 的$N$个 AuNPs 分散在体积为$V$、温度为$T$和介电常数为$\varepsilon_r$的溶剂中。DNA 复制的延伸阶段在 323 K 的恒定温度下进行。模拟体系的粒子数目$N$可由公式$N = C \cdot V \cdot N_A$计算，其中$N_A$是阿伏伽德罗常数，$C$代表 AuNPs 的浓度。将平均电荷作为可调参数，不同粒子的带电量遵循泊松分布。随着平均电荷数从较小值增加到几百个，径向分布函数逐渐呈现明显的峰值。

具体过程概括如下：①从系统中随机选择一个粒子，设其位置为$r_i = (x_i, y_i, z_i)$，计算该粒子与其他粒子之间的相互作用能量：

$$u(r_i) = \sum_{j \neq i} \frac{q_i q_j}{4\pi\varepsilon_0 \varepsilon_r r_{ij}} \tag{1}$$

其中$q_i$和$q_j$是粒子的电荷，$r_{ij} = |r_i - r_j|$。②在三维空间中随机移动到一个新的位置$r'_i = (x'_i, y'_i, z'_i)$，计算该位置的系统能量$u(r'_i)$。③确定能量差$\Delta u = u(r'_i) - u(r_i)$。若$\Delta u < 0$，接受新位置；若$\Delta u > 0$，接受概率为$p = \exp(-\Delta u / k_B T)$，其中$k_B$为玻尔兹曼常数，即产生一个介于 0 和 1 之间的随机数$\delta$，若$\delta < p$，则可接受新状态。重复该过程直至系统达到平衡。

实验使用公式$d = 1/\sqrt[3]{N_A C}$描述粒子间距，并确定了自由基分布峰的峰值位置$D$（$D = 1.9d$）[14, 15]。通过分析实验数据，我们由 AuNP 的粒子间距$d$和$D$推导出相应的浓度。改变粒子的带电量$q$，得到不同浓度下 AuNP 的径向分布函数，

如图 1 所示。结果显示，随着电荷的增加，图 1（c）中的粒子分布展现出准周期性，具有明显的峰值和振荡，且径向分布特征峰值与实验[14]中预测的粒子间距相近。这归因于胶体粒子之间静电斥力的增加，粒子排布变得有序。

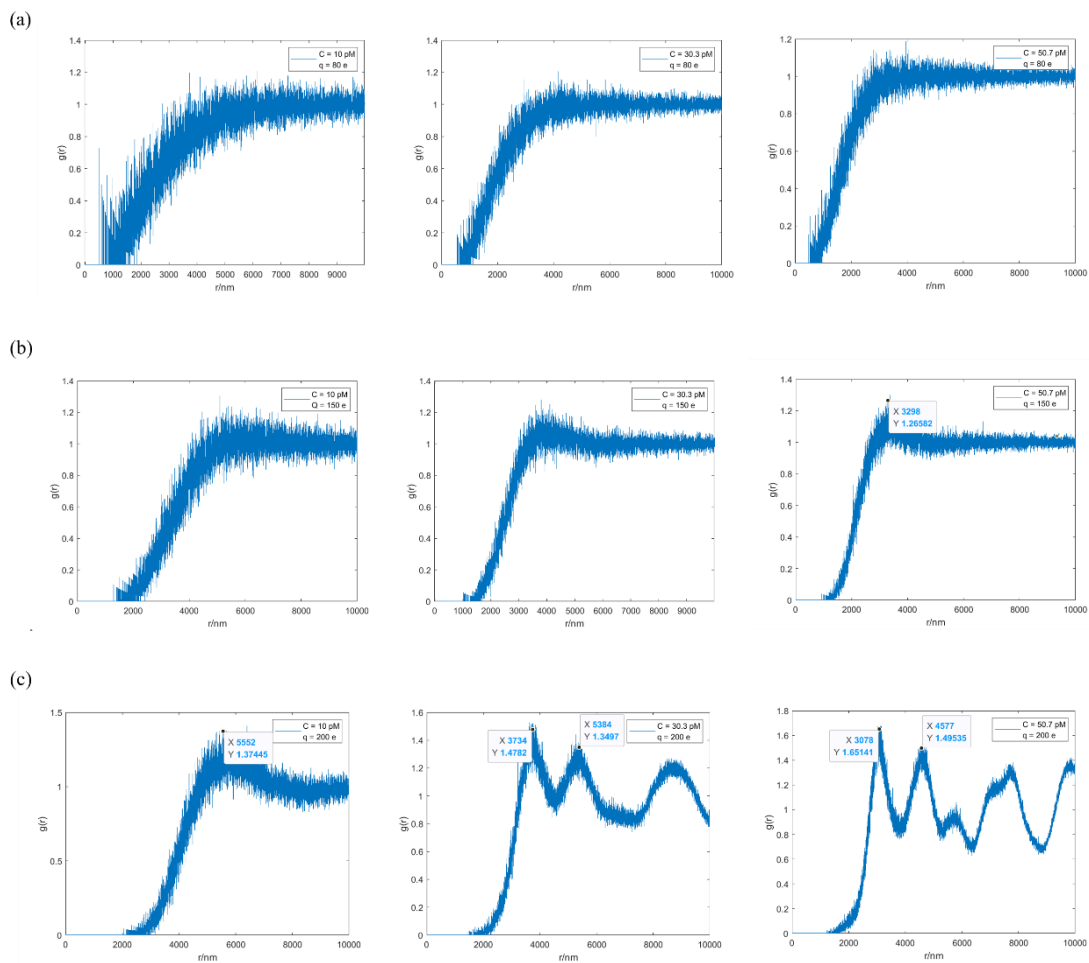

图 1 不同浓度 AuNP（10 pM；30.3 pM；50.7 pM）带电量分别为 80 e（图 a），150 e（图 b），200 e（图 c）的径向分布函数

Fig. 1 The radial distribution functions of colloidal AuNPs at different concentrations (10 pM, 30.3 pM, 50.7 pM) with charges of 80 e (Figure a), 150 e (Figure b), and 200 e (Figure c), respectively

2.2 相干瑞利散射

在胶体溶液中，AuNP 的近周期排列为其光学特性奠定了基础。根据参考文献[14]和[15]中的计算，水解反应释放的光子波长为 9.2 μm，溶液折射率约为 1.3，相应波长变为 7 μm。理解生物光子与金属纳米粒子的相互作用如何提升实验效率成为一个关键问题。在 PCR 过程中， dNTP 水解释放的能量部分以光

子形式发射出来。该过程有效地为实验环境提供了一个稳定且固定频率的入射光子流。在稀溶液中，胶体粒子间距较大且 AuNPs 半径远小于光子波长，因此会发生瑞利散射。所产生的散射波将以球面波的形式呈现，如图 2 所示。

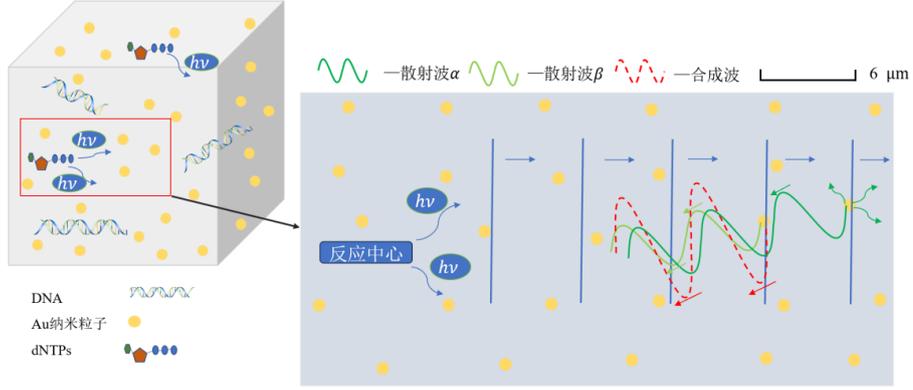

图 2 AuNP 散射平面波示意图。当平面波沿着正向传播并与粒子相互作用时，会发生散射。粒子的准周期分布增加了相邻粒子之间发生散射的可能性。当粒子的光程差与入射光子波长的半整数倍匹配时，散射波相位一致。这种相干叠加会显著增加能量在特定方向上的传播

Fig. 2 Illustration of scattering of a plane wave by Au nanoparticles. When a plane wave propagates in the positive direction and interacts with particles, scattering occurs. The quasi-periodic distribution of particles increases the likelihood of scattering events between adjacent particles. When the optical path difference of a particle matches integer multiples of half the wavelength of the incident photon, the phase of the scattered wave is consistent. This coherent superposition significantly enhances energy propagation in a specific direction.

当入射电场与 AuNP 相互作用时，它在纳米粒子中诱导了一个随时间变化的偶极矩。偶极子以光源的形式向外场辐射出散射光，偶极矩由方程（2）确定：

$$\boldsymbol{P} = \gamma \boldsymbol{E}_i \qquad (2)$$

其中 $\boldsymbol{E}_i$ 为入射电场，$\gamma$ 为 AuNP 的极化率。不失一般性，设入射波沿 $\hat{x}$ 方向，$\boldsymbol{E}_i$ 沿 $\hat{z}$ 方向。选取原点作为参考点，则位于 $\vec{r}$ 点的 AuNP 在该处产生的散射振幅为：

$$\boldsymbol{E} \propto P \sin(\theta) \frac{1}{r} e^{-ikr} \boldsymbol{e}_\theta \qquad (3)$$

其中 $\boldsymbol{e}_\theta$ 为极坐标的单位基矢，$k$ 是波数。考虑两个邻近粒子散射波的相干叠加：

$$\boldsymbol{E}_1 \propto \sin(\theta_1) \frac{1}{r_1} e^{i\delta_1} \boldsymbol{e}_{\theta_1} \qquad (4a)$$

$$\boldsymbol{E}_2 \propto \sin(\theta_2) \frac{1}{r_2} e^{i\delta_2} \boldsymbol{e}_{\theta_2} \qquad (4b)$$

其中$\delta_1$，$\delta_2$分别代表两列波的相位。相干光的电场为：

$$\boldsymbol{E}_{tot} = \boldsymbol{E}_1 + \boldsymbol{E}_2 \tag{5}$$

得到相干光的能量强度为：

$$I_{tot} \propto \left| Re\left[\frac{1}{r_1}\sin(\theta_1)\boldsymbol{e}_{\theta_1} + \frac{1}{r_2}\sin(\theta_2)\boldsymbol{e}_{\theta_2}e^{i(\delta_2-\delta_1)}\right]\right|^2 \tag{6}$$

我们针对上述相干过程使用 Matlab 进行了蒙特卡洛模拟。将固定数量的粒子（$N = 1500$）限制在边长为$2L$的立方体盒内，模拟 AuNP 在 AuNP-qPCR 实验中的环境。$L$的数值由 AuNPs 的粒子数$N$与其浓度$C$之间的关系确定，即$2L = \sqrt[3]{N/(C \cdot N_a)}$。引入周期性边界条件以确保离开模拟边界的粒子将从对面重新进入，避免结果中与边界相关的偏差。

模拟以 AuNPs 在胶体溶液中的均匀分布开始，使用第 2.1 节中概述的标准算法在系统中建立胶体粒子的平衡分布。电场方向(0,0,1)由入射平面波(1,0,0)推导得来。随机选取一个参考粒子，并确定其最近的相邻粒子，考虑这两个粒子的相干散射波叠加。散射波的矢量表达式在公式（3）中给出其振幅和相位。计算出两个粒子离原点(0,0,0)的距离，可认为其近似相等。因此，由公式（4）推导出两列散射波的电场表达式，通过对矢量和取平方得到原点处散射波的相对光强。其中，在半径 r 附近找到粒子的概率与$r^2$成正比关系，该因子消除公式（6）中的$(1/r)^2$。模拟使用蒙特卡洛方法进行了$10^6$次迭代，每$10^4$帧计算一次光强。每次计算随机选取 1000 对粒子，对结果进行平均以确保稳定性。随后，将模拟得到的相对光强值与实验数据中的效率提升情况进行比较，评估该粗略化方法的可行性。

# 3 结果与分析

3.1 数据拟合

参考 qPCR 扩增效率周期性变化的实验数据[15]，筛选与实验相同的 x 轴数值，绘制出 AuNPs 在低浓度下（2~20.9 pM）的粒子间距与散射波干涉后相对光强之间的关系图。图 3 描述了整体趋势，与实验结果密切吻合。

对光强数据进行多项式拟合分析，揭示了相对光强中明显的峰谷。这表征实验效率的增幅，并直观呈现周期性振荡。重要的是，周期与生物光子波长一致。表明通过调节 AuNPs 浓度以改变粒子间距，影响散射波的干涉模式，进而影响实验效率。瑞利散射效应本身较弱，但 PCR 过程的迭代性质使得信号呈指数级扩增。在每个 PCR 周期中，生成的产物作为后续过程的底物，将最初的微弱信号放大到可检测和可分析的水平。这为 AuNP-qPCR 实验中观察到的效率波动提供了可靠的解释。

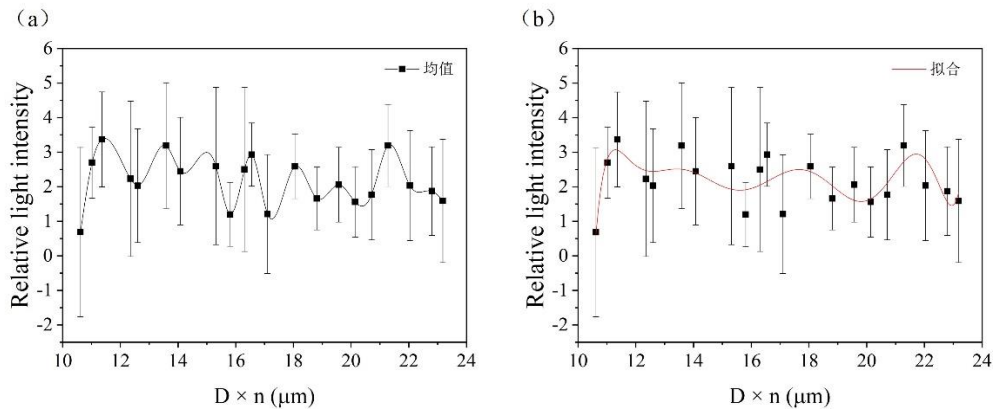

图 3 相对光强在特定参考点上随直径 15 nm AuNPs 之间距离 $D$ 的周期性振荡以及光强值的多项式拟合结果。（a）选择 $D$ 的方法基于实验[14]中的描述，其中 $n \approx 1.3$，为溶液的折射率，$D = 1.9d$，便于进行比较。（b）光强值的多项式拟合显示出与观测到的实验趋势吻合的准周期性振荡

Fig.3 The relative light intensity oscillates with the distance $D$ between 15 nm AuNPs at a specific reference point and the polynomial fitting results of the light intensity value. (a) The method of selecting $D$ is based on the description in the experiment [14], where $n \approx 1.3$, representing the refractive index of the solution, and $D = 1.9d$, for ease of comparison. (b) Polynomial fitting of the light intensity values shows quasi-periodic oscillations consistent with the observed experimental trend.

## 3.2 AuNP 的量子行为

### 3.2.1 AuNP 费米能近邻能级分立度

瑞利散射模型简洁地再现了实验中观察到的效率周期性振荡现象。此外，我们探究了纳米材料独特的量子尺寸效应对增强实验效率的潜在贡献。在这项实验中，AuNPs 的存在是一个关键因素，需要对其费米能级近邻的离散能级进行定量分析，并探索导电电子与生物光子之间的相互作用。为了保持简单性和普适性，假设每个 AuNP 贡献一个电子，粒子的费米能$E_F$为[23, 24]

$$E_F = \frac{\hbar^2 k_F^2}{2m} \quad (7a)$$

$$k_F = \left(\frac{3\pi^2 \rho N_A}{M}\right)^{\frac{1}{3}} \quad (7b)$$

其中$k_F$是费米波矢，$\hbar$是约化普朗克常数，$\rho = 19.32 \text{ g/cm}^3$，表示金元素的密度；$m = 9.11 \times 10^{(-31)}$ Kg，表示电子质量；$M = 196.97$ g/mol，表示金的摩尔质量。AuNP 的费米能量确定为$E_F = 11.42$ eV。

将粒子内部的电子运动看作一个简谐振子。费米能级的方程为$E_F = \frac{1}{2}m\omega^2 R^2$，其中$R$表示 AuNP 的半径，角频率确定为$\omega = 2.67 \times 10^{14}$ s$^{-1}$。根据三维谐振子模型，本征能量为[25]

$$E_n = \left(n_x + n_y + n_z + \frac{3}{2}\right)\hbar\omega = \left(n + \frac{3}{2}\right)\hbar\omega \quad (8)$$

其中$n$为量子数。能级差$\Delta E = \hbar\omega = 0.18$ eV。相应的波长计算公式为$\lambda = 2\pi c/\omega$，其中$c$为真空光速，$\lambda = 7.06$ μm。这只是一个粗略的近似。采用另一种模型为估算提供参考，把费米能级近似为一个无限深的方势阱。此时，本征能级为[25]

$$E_n = \frac{\hbar^2 k^2}{2m} = \frac{\hbar^2 (k_x^2 + k_y^2 + k_z^2)}{2m} \quad (9)$$

其中$k_x = n\pi/(2R)$，可得$k_F = \sqrt{2mE_F/\hbar^2} = 1.74 \times 10^{10}$，取微元$\delta k_x = \pi/(2R)$，

计算能量差 $\delta E = \pi E_F/(\sqrt{3}R k_F)$，代入数值，得到 $\delta E = 0.16$ eV。对应波长用 $E = hc/\lambda$ 确定，为 $\lambda = 7.78$ μm。

使用三维谐振子模型和无限深方势阱模型近似描述了 AuNP 内部电子的特性。在谐振子模型中，确定了固定的能级间距为 7.06 μm。在方势阱模型中，能级间距与波数成正比，在费米能附近计算得到的能级宽度为 7.78 μm。这两个模型结果的一致性确认了估算的准确性。

### 3.2.2 AuNP 内部的电子态

确定 AuNP 费米能级附近能级的宽度与生物光子波长相吻合后，利用玻尔兹曼常数 $k_B$ 与温度 $T$（保持在恒定 323 $K$）相乘以估算 AuNP 内电子的热能[26]，得到 $E = 0.028$ eV。这一数值表明电子处于相对较低的激发态，暗示了其吸收光子并在光子相互作用期间释放出更高能量的潜力。这为阐明文献[15]中讨论的光子反应之间的能量差异提供了一个新的视角，即 dNTPs 水解反应的子反应 I 和 IV 之间的能量缺陷如何通过热能来补偿。实质上，水解反应的下游过程释放出光子，但对于上游过程所需的光子存在频率差异，可通过在吸收光子的共振过程中固有于电子态中的热能降低能量势垒。

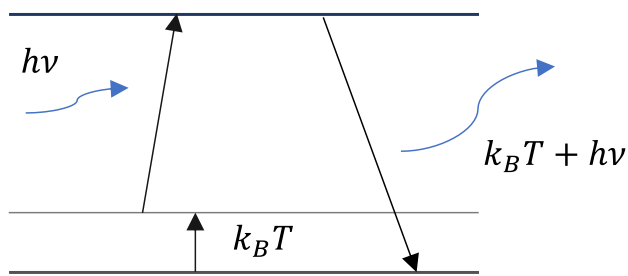

图 4 光子与 AuNP 内部电子的相互作用示意图。系统中的电子具有一定热能，当光子能量与其能级宽度匹配时，电子通过共振吸收光子进入激发态，随后放出能量回到基态

Fig. 4 Illustration of interaction between photons and electrons in an Au nanoparticle. The electrons in the system possess thermal energy. When the energy of a photon matches the energy level width of an electron, the electron resonantly absorbs the photon and transitions to an excited state. Subsequently, it releases energy and returns to the ground state.

### 3.3 储存能量的再利用

如图 4 所示，讨论 AuNPs 的电子跃迁过程如何影响后续实验是至关重要的。当电子从激发态返回基态时，能量通过辐射和非辐射机制释放。主要机制是自发辐射[27]，即 AuNPs 在这一过渡过程中发射光子，这些光子被反应中心 dNTPs 吸收，推动反应向前发展。另一个重要的非辐射过程是 Förster 共振能量转移（Förster resonance energy transfer, FRET）[28]，通过纳米级别的偶极子相互作用将 AuNPs 的能量转移到附近的分子。在光子波长整数倍处的共振激子极化子进行的非辐射能量转移也发挥一定作用[29]。在本实验中，该机制的有效性有限。此外，AuNPs 还可以通过与相邻分子或粒子的碰撞过程传递能量[30]。这些机制共同影响着 AuNP-qPCR 实验中反应的进展和效率。

## 4 讨论

在浓度较高的 AuNPs 存在的情况下，单链 DNA 结合蛋白（single-stranded DNA binding protein, SSB）诱导的增强效应可能会影响光学效应的确定[15]，因此，本研究主要关注低浓度下 AuNP 的光学行为。在此区域 AuNP 的增强效应超过了 SSB[13]，显示其背后机理尚待研究，因此我们利用 AuNP 的瑞利散射相干模型可以为探明其机制提供一个新方向。

本研究利用基于长程库仑力的蒙特卡洛模拟探究了 AuNPs 的行为。AuNPs 的有效电荷（$q$）遵循高斯定理[31, 32]，可由方程$q = 4\pi\varepsilon rU$确定，其中$\varepsilon$表示溶剂的介电常数（$\varepsilon = \varepsilon_0 \cdot \varepsilon_r$，$\varepsilon_r$为水的相对介电常数），$r$是粒子的等效半径。值得注意的是，单个粒子表面电位可保持恒定，但 AuNPs 的电荷不是一个固定值，受到运动速度和吸附在表面上的颗粒总数等各种因素的影响。粒子之间的相互作用通过裸电荷之间的库仑相互作用来展现，主要与其周围扩散层的速度有关。

该模拟采用了相干瑞利散射模型评估参考点的相对光强，将 AuNPs 的平均电荷设置在 200e 到 250e 的范围内。研究发现，当 AuNPs 达到相对平衡状态时（即出现径向分布峰），PCR 效率的振荡现象在不同程度上得以重现。表明，在所研究范围内，有效电荷的具体数值对振荡现象的影响可忽略。随着未来研究中检测技术的进步和实验数据的积累，本模型的精度和准确性将进一步提高。这将深化我们对 AuNP 动态行为及其在各种应用中意义的理解。

综上，本研究利用蒙特卡洛技术，探究了低浓度下胶体体系中 AuNPs 的分布模型。其主要目标是理解 AuNP-qPCR 实验效率的周期性振荡现象。通过调节粒子的有效电荷，研究建立了一个在平衡态下具有有序分布的颗粒模型。利用该模型计算了经过颗粒瑞利散射后颗粒距离与相干光相对强度之间的关系，揭示了 AuNP 浓度与 qPCR 实验效率之间周期性相关的光量子机制。尽管散射效应较弱，PCR 迭代特性可将其放大到可测量的水平，从而验证了我们的理论解释。其次，探讨了 AuNP 的量子效应，观察到在费米能级附近的能级分裂，与生物光子保持一致。这一观察结果提示了粒子对光子的选择性共振吸收，并在返回基态时发射更高能量的光子，为克服水解反应中的势能屏障提供所需能量。这为阐明实验中观察到的光子频率差异及其与热能的关系提供了新的角度。最后，第 3.3 节讨论了进一步利用能量的潜在途径，为未来深入研究提供了参考。这些发现对于解释 AuNP-qPCR 实验效率的周期性振荡现象具有重要意义。

此外，PCR 凭借其扩增效应，有望成为一个精确的探测手段，用于研究和探索更多生物量子现象，为此类研究提供工具和平台。这反过来将深化研究者们对生物体中量子行为的理解和揭示，推动生物学和量子生物学领域的发展和进步。

参考文献

[1] Saiki R K, Scharf S, Faloona F, Mullis K B, Horn G T, Erlich H A, Arnheim N 1985 *Science* **230** 1350

[2] SAiK R K, GELFAND D H, STOFFEL S, SCHARF S J, HIGUCHI R, HoRN G T, MULLIS K B, ERLICH H A 1988 *Science* **239** 487

[3] NAJAFOV A, HOXHAJ G 2016 *PCR Guru: An Ultimate Benchtop Reference for Molecular Biologists* (Cambridge, Massachusetts: Academic Press) pp1-16

[4] Mullis K B 1990 *Sci. Am.* **262** 56

[5] Blumenfeld N R, Bolene M A E, Jaspan M, Ayers A G, Zarrandikoetxea S, Freudman J, Shah N, Tolwani A M, Hu Y H, Chern T L, Rogot J, Behnam V, Sekhar A, Liu X Y, Onalir B, Kasumi R, Sanogo A, Human K, Murakami K, Totapally G S, Fasciano M, Sia S K 2022 *Nat. Nanotechnol* **17** 984

[6] Michalet X, Pinaud F F, Bentolila L A, Tsay J M, Doose S, Li J J, Sundaresan G, Wu A M, Gambhir S S, Weiss S 2005 *Science* **307** 538

[7] Tabatabaei M S, Islam R, Ahmed M 2021 *Anal. Chim. Acta* **1143** 250

[8] Li M, Lin Y C, Wu C C, Liu H S 2005 *Nucleic Acids Res.* **33** e184

[9] Liang Y, Luo F, Lin Y, Zhou Q, Jiang G 2009 *Carbon* **47** 1457

[10] Petralia S, Barbuzzi T, Ventimiglia G 2012 *Materials Science and Engineering: C* **32** 848

[11] Abdul Khaliq R, Kafafy R, Salleh H M, Faris W F 2012 *Nanotechnology* **23** 455106

[12] Wang L, Zhu Y, Jiang Y, Qiao R, Zhu S, Chen W, Xu C 2009 *J. Phys. Chem. B*

**113** 7637

[13] Li H K, Huang J H, Lv J H, An H J, Zhang X D, Zhang Z Z, Fan C H, Hu J 2005 *Angew. Chem. Int. Ed.* **44** 5100

[14] Li N, Peng D, Zhang X, Shu Y, Zhang F, Jiang L, Song B 2020 *Nano Res.* **14** 40

[15] Yang Y, Peng D, Gu Z, Jiang L, Song B 2023 *Chemistry* **29** e202203513

[16] Naraginti S, Li Y 2017 *J. Photochem. Photobiol. B, Biol.* **170** 225

[17] Singh P, Pandit S, Garnaes J, Tunjic S, Mokkapati V R, Sultan A, Thygesen A, Mackevica A, Mateiu R V, Daugaard A E, Baun A, Mijakovic I 2018 *Int. J. Nanomedicine* **13** 3571

[18] Honary S, Zahir F 2013 *Trop. J. Pharm. Res.* **12** 255

[19] Fernandez-Nieves A, Puertas A M 2016 *Fluids, colloids and soft materials: an introduction to soft matter physics* (John Wiley & Sons)

[20] Cheng R T, Zhu X M, Yuan R S, Ao P 2018 *Proceedings of the 37th Chinese Control Conference* Wu Han, July 25-27, 2018 p8221

[21] Kirkpatrick S, Gelatt C D, Jr., Vecchi M P 1983 *Science* **220** 671

[22] Frenkel D, Smit B 2001 *Understanding Molecular Simulation: From Algorithms to Applications* (Second Edition) (London: Academic Press) pp23-48

[23] Kittel C 2005 *Introduction to solid state physics* (Eighth Edition) (New Jersey: John Wiley & Sons) pp6.1-6.4

[24] Ashcroft N W, Mermin N D 1976 *Solid State Physics* (New York: Holt, Rinehart and Winston) pp589-615

[25] Zeng J Y 2007 *Quantum Mechanics* (Fourth Edition) (Beijing: Science Press)

pp60-419 (in Chinese) [曾瑾言 2007 量子力学 （第四版）（北京：科学出版社）第 60—419 页]

[26] Feynman R P, Leighton R B, M Sands（translated by Zheng Y L, Hua H M, Wu Z Y） 2005 *The Feynman's Lectures on Physics* (Shanghai: Shanghai Scientific & Technical Publishers) pp411-415 (in Chinese）[费恩曼, 莱顿, 莱兹著 （郑永令，华宏鸣，吴子仪等译） 2005 费恩曼物理学讲义（上海：上海科学技术出版社）第 411—415 页] {译著}

[27] Vahala K J 2003 *Nature* **424** 839

[28] Clegg R M 2006 *The history of FRET: from conception through the labors of birth* (Boston, MA: Springer) pp1-45

[29] Chen Y C, Song B, Leggett A J, Ao P, Zhu X M 2019 *Phys. Rev. Lett.* **122** 257402

[30] Blauer J A, Solomon W C, Sentman L H, Owens T W 1972 *J. Chem. Phys.* **57** 3277

[31] Quesada-Pérez M, Martín-Molina A 2013 *Soft Matter* **9** 7086

[32] Wu J Z 2022 *Chem. Rev.* **122** 10821

# The light quantum mechanism of PCR efficiency oscillation with gold nanoparticle concentration[*]

Fang Huan-Huan[1)] Chen Yong-Cong[1)†] Liu Ze-Fei[1)] Zhu Xiao-Mei[2)] Ao Ping[3)]

1) (Shanghai Center for Quantitative Life Sciences and Physics Department,

Shanghai University, Shanghai 200444, China)

2) Shanghai Key Lab of Modern Optical System, School of Optical-Electrical Computer

Engineering, University of Shanghai for Science and Technology, Shanghai 200093, China)

3) College of Biomedical Engineering, Sichuan University, Chengdu 610065, China)

Abstract

The widespread application of nanomaterials in polymerase chain reaction (PCR) technology has opened new avenues for improving detection methods in the biomedical field. Recent experiments (Chem. Eur. J. 2023, e202203513) have revealed oscillatory behavior between PCR efficiency and the concentration of gold nanoparticles in the pM range, potentially linked to the long-range Coulomb interactions among charged colloidal particles and the quantum size effect of nanoparticle electronic states. Through Monte Carlo simulation, we discovered that the radial distribution function of gold nanoparticles in solution gradually exhibits peak characteristics with increasing charge, triggering coherent photon behavior in Rayleigh scattering within the solution, thereby influencing the efficiency of reusing released photons in the PCR chain reaction. The study demonstrates that the oscillation period aligns with the wavelength of downstream reaction photons, while their energy matches the width of energy levels near the Fermi level of gold nanoparticles. The latter can absorb and store electron states internally, promoting upstream PCR reactions through subsequent re-release, and compensating for energy deficiencies through the Boltzmann distribution of electrons. This work is poised to advance the application of PCR-specific precise detection methods in the field of quantum biotechnology.

**Keywords:** gold nanoparticles, Polymerase Chain Reaction, long-range Coulomb action, Rayleigh coherent scattering, quantum size effect

**PACS:** 42.25.Kb, 61.46.Df, 71.18.+y, 78.35.+c

\* Supported in part by the National Natural Science Foundation of China (Grant No. 16Z103060007).

† Corresponding author. E-mail: chenyongcong@shu.edu.cn

   First author. E-mail: 2229537512@shu.edu.cn